\documentclass[11pt,oneside,a4paper]{article}
\usepackage{url} 
\usepackage{appendix}
\usepackage[british]{babel}
\usepackage[hmargin=2.5 cm,vmargin=2.5 cm]{geometry} 
\usepackage{amsmath}
\usepackage{booktabs} 
\usepackage[pdftex]{graphicx}
\usepackage[final]{pdfpages}  
\usepackage{float}
\usepackage{amssymb}
\usepackage{tikz}
\usepackage{pgfplots}
\usepackage{listings}
\usepackage{color}
\usepackage{tocbibind}
\usepackage{pdflscape}
\usepackage{pdfpages}
\usepackage[scientific-notation=true]{siunitx}
\usepackage{hyperref}
\usepackage{verbatim}
\usepackage{subcaption}
\usepackage{setspace}
\usepackage[ruled, linesnumbered]{algorithm2e}
\usepackage{nomencl}
\usepackage{sectsty}
\makenomenclature
\makeindex

\DeclareMathAlphabet{\mathpzc}{OT1}{pzc}{m}{it} 

\definecolor{mygreen}{RGB}{28,172,0} 
\definecolor{mylilas}{RGB}{170,55,241} 
\DeclareMathAlphabet{\mathpzc}{OT1}{pzc}{m}{it} 

\begin{document}
\onehalfspacing

\begin{center}
\vspace{4cm}
{\Large \bf A computational study of viscoelastic blood flow}\\
{\Large \bf in an arteriovenous fistula}\\

\bigskip

N. Vundla$^{1,2}$  and B.D. Reddy$^{1,3}$ \\ 

$^1$ Centre for Research in Computational and Applied Mechanics \\ 
$^2$ Department of Mechanical Engineering \\
${^3}$ Department of Mathematics and Applied Mathematics \\
University of Cape Town, 7701 Rondebosch, South Africa 
\end{center}

\section*{Abstract}

A finite element analysis of flows of an Oldroyd-B fluid is developed, to simulate blood flow in an arteriovenous fistula. The model uses a combination of a standard conforming finite element approximation for the momentum equation, and the discontinuous Galerkin method, with upwinding, for the equation governing the evolution of the extra stress. The model is verified for a range of values of Weissenberg number We by applying it to the benchmark problem of flow past a cylinder in a channel. The main application is to flow in an arteriovenous fistula, the geometry of which is based on patient-specific data. Results for Oldroyd-B fluids are compared with those for a Newtonian fluid as well as with data from patient-specific velocity MRI scans. Features such as streamlines and regions of recirculation are similar across a range of values of We and the Newtonian case. There is however a strong dependence of maximum wall shear stress on We, with values for the viscoelastic fluid in all cases being higher than that for the Newtonian case.

\section{Introduction} 


While the simple model of a Newtonian fluid suffices for a wide range of fluid behaviour, there are many fluids whose behaviour cannot be satisfactorily modelled as Newtonian. Extensions, for example to non-Newtonian models, then become necessary. Examples of such models, applicable to a range of fluids, include the Upper Convected Maxwell, Oldroyd and Oldroyd-B, Phan-Thien-Tanner (PTT), Finite Extendible Nonlinear Elasticity (FENE) and Giesekus models \cite{Owens-Phillips2002}.\\

The Oldroyd-B model, which is adopted in this work, has been used in applications such as heating, blood flow and flow through porous media \cite{Owens-Phillips2002}. In \cite{YELESWARAPU1998257} a generalized form of the model was used to simulate blood flow, with experimental validation using the results of tests on porcine blood. Similar work, with the emphasis on shear thinning, was presented in \cite{Donev2014}. \\

Various numerical approaches to the equations for Oldroyd-B fluids have been investigated: these include Discrete Elastic Viscous Stress Splitting (DEVSS) \cite{Kim2004}, Local Projection Stabilization \cite{Venkatesan2017}, Galerkin Least Squares (GLS) \cite{Coronado2006}, and the extended finite element method (XFEM) \cite{Choi2012}. In the majority of these studies a major challenge has been that of obtaining convergent results at higher Weissenberg numbers \cite{Hulsen2005}. \\

Blood comprises a suspension of red and white blood cells and platelets in a plasma [1]. In larger arteries, blood has been traditionally modelled as an incompressible Newtonian fluid with good experimental correlation (see [2]). However, in smaller channels blood exhibits non-Newtonian properties such as shear-thinning as well as viscoelasticity. \\

Models that account for both shear-thinning and viscoelastic effects include that presented in [4], where an empirically fitted viscosity function was incorporated into a generalized Oldroyd-B model, and experimentally validated against an in vitro experiment with porcine blood.
Other numerical studies based on this model include [5, 6, 7], with different viscosity functions having been similarly incorporated into the generalized Oldroyd-B model. \\

The focus of this work is on a computational study of viscoelastic flows in a complex domain whose geometry derives from a patient-specific  arteriovenous fistula. The aim of the study is to extend an earlier biomechanical investigation, reported in \cite{DeVilliers2017}, of blood flow in such a domain. In that study the fluid was modelled as Newtonian, and the vessel walls treated as deformable. Similar studies have been reported in 
\cite{Guess2016,Vignon-Clementel2010}. The objective of this work is to investigate features with the adoption of an Oldroyd-B model, and to compare these with results obtained with the assumption of Newtonian flow. \\

A constant viscosity Oldroyd-B model is adopted, and the resulting set of governing equations solved approximately using a combination of conforming and discontinuous Galerkin finite element methods. The latter is used for the constitutive relation involving the extra stress, as it is well suited to developing stable approximations for equations of advection-diffusion type, such as that considered here. \\

The structure of the remainder of this work is as follows. The governing equations of the problem and details of the discretization schemes used are presented in Section 2. In Section 3 the approximation procedure is applied to the benchmark problem of flow around a cylinder. The main example, of flow in an arteriovenous fistula, is presented in Section 4, with comparisons given between results for the case of Newtonian and Oldroyd-B fluids, and also of patient-specific MRI data. \\

\section{Governing equations and their discretization}
\subsection{Governing equations}
Consider a fluid occupying a domain $\Omega \subset \mathbb{R}^d$ ($d = 2, 3$) with boundary $\partial \Omega$. The governing equations for flows of a generalized Oldroyd B fluid are, in dimensionless form, and in the absence of a body force \cite{Owens-Phillips2002},
\begin{subequations}
\begin{align}
& \text{Re} \left(\dfrac{\partial{\boldsymbol  u}}{\partial t} +({\boldsymbol  u} \cdot {\boldsymbol \nabla}){\boldsymbol  u} \right) -\mbox{div}\,{\boldsymbol T}
= {\boldsymbol 0}, \\
& {\boldsymbol T} = -p\boldsymbol{I} + \beta(\nabla \boldsymbol{u} + (\nabla \boldsymbol{u})^T) + \boldsymbol{\tau},   \\
& {\boldsymbol \nabla} \cdot{\boldsymbol  u} = 0,   \\
& {\boldsymbol \tau} + \text{We} \stackrel{{\nabla}}{{\boldsymbol \tau}} - 2(1 -\beta) {\boldsymbol D} = {\boldsymbol 0}.
\end{align}
\label{eqn:oldb_strong}
\end{subequations}
Here ${\boldsymbol u}$ denotes the velocity, ${\boldsymbol T}$ the Cauchy stress, $p$ the pressure, and ${\boldsymbol \tau}$ the extra stress; the rate of deformation tensor $\boldsymbol D$ is given by
\begin{align*}
\boldsymbol D &= \dfrac{1}{2}({\boldsymbol \nabla}{\boldsymbol  u} + ({\boldsymbol \nabla}{\boldsymbol  u})^T) .
\end{align*}
The upper convective derivative $\stackrel{{\nabla}}{{\boldsymbol \tau}}$ of the stress ${\boldsymbol \tau}$ is defined by
\begin{align}
\stackrel{{\nabla}}{{\boldsymbol \tau}} =\left(\dfrac{\partial{\boldsymbol  u}}{\partial t} +({\boldsymbol  u} \cdot {\boldsymbol \nabla}){\boldsymbol \tau} \right) - ({\boldsymbol \nabla}{\boldsymbol  u} ){\boldsymbol \tau} - {\boldsymbol \tau}({\boldsymbol \nabla}{\boldsymbol  u})^{T}.
\label{eqn:upper_conv}
\end{align}
The quantity $\beta = \eta_s / \eta$ is the ratio of the polymeric viscosity $\eta_s$ to total viscosity, $\eta$, which is the sum of solvent and polymer parts $\eta_s$ and $\eta_p$. Nondimensionalization is achieved with the introduction of a characteristic length $L$ and velocity $U$, and the dimensionless parameters are the Reynolds number $\text{Re}$ and Weissenberg nmber $\text{We}$, defined by
\begin{equation}
\text{Re} = \frac{\rho UL}{\eta},\qquad \text{We} = \frac{\lambda_1}{LU}.
\label{eqn:ReWe}
\end{equation} 
In the definition of Re, $\rho$ is the mass density, while in the definition of We, $\lambda_1$ is the viscoelastic relaxation time. This parameter is thus the ratio of viscoelastic to viscous quantities, and serves as a measure of the degree of viscoelasticity.\\

The boundary $\Gamma$ has outward unit normal $\boldsymbol n$ and is subdivided into two non-overlapping parts $\Gamma_D$ and $\Gamma_N$ referred to as the Dirichlet (essential) and Neumann (natural) boundaries, with $\Gamma_D \cup \Gamma_N = \Gamma$. The inflow boundary, that is, that portion of $\Gamma$ along which $\boldsymbol u \cdot\boldsymbol n < 0$, is denoted by $\Gamma_i$. Boundary conditions are prescribed as follows:
\begin{subequations}
\begin{align}
\boldsymbol u = \boldsymbol g \text{ on } \Gamma_D ,   \\ 
\boldsymbol \tau = \boldsymbol {\bar \tau} \text{ on } \Gamma_i ,   \\
  {\boldsymbol \sigma}{\boldsymbol n}  = \boldsymbol {\bar{t}} \text{ on } \Gamma_N .
\label{eqn:poisson_str}
\end{align}
\end{subequations}
In addition, the initial conditions are 
\begin{subequations}
\begin{align}
\boldsymbol u (\boldsymbol x,0) = \boldsymbol 0, \\
\boldsymbol \tau (\boldsymbol x,0) = \boldsymbol 0. 
\end{align}
\end{subequations}

%


\subsection{Discretization}

{\bf Time-discretization.}\ \ 
A backward Euler scheme is used to discretize in time, so that the resulting equations are, at time step $n$, 
\begin{subequations} 
\begin{align}
    \dfrac{\text {Re} }{\Delta t} \boldsymbol u^n +  \text {Re} ({\boldsymbol u}\cdot{\nabla})\boldsymbol u^n + {\nabla p^n} - \beta \nabla ^2 {\boldsymbol u^n} - \nabla \cdot {\boldsymbol \tau^n } &= \dfrac{\text {Re} }{\Delta t} \boldsymbol u^{n-1},\\
    \nabla \cdot {\boldsymbol u^n } &= 0, \\
    \left(1 + \dfrac{\text{We} }{\Delta t}\right){\boldsymbol \tau^n } + \text{We} ({\boldsymbol u}\cdot{ \nabla) \boldsymbol \tau^n } - ({\nabla \boldsymbol u^n}){\boldsymbol \tau^n } - {\boldsymbol \tau^n } ({\nabla \boldsymbol u^n})^T & \nonumber \\ - (1 -\beta)({\nabla \boldsymbol u^n} + ({\nabla \boldsymbol u^n})^T) &= \dfrac{\text{We} }{\Delta t}{\boldsymbol \tau^{n - 1} }\, .
    \end{align} 
    \label{eqn:b_strong}
    \end{subequations}

{\bf Spatial discretization.}\ \ 
We begin by formulating the governing equations (\ref{eqn:b_strong})  in weak form, by taking the inner product respectively with test functions $\boldsymbol w$, $q$ and $\boldsymbol \sigma$, integrating, and integrating by parts where relevant. This gives the set of equations
\begin{subequations}
    \begin{align}
    &\int_{\Omega}\dfrac{\text {Re} }{\Delta t} \boldsymbol u^n \cdot \boldsymbol w\ dV +  \int_{\Omega}\text {Re} ({\boldsymbol u}^n\cdot\nabla  ){\boldsymbol u^n} \cdot {\boldsymbol w}\ dV   + \int_{\Omega}\left[ \beta \nabla  {\boldsymbol u^n} \cdot \nabla   {\boldsymbol w} + {\boldsymbol \tau^n } \cdot \nabla {\boldsymbol w}\right]\ dV \nonumber \\ 
    \qquad & - \int_{\Omega}{p^n} (\nabla \cdot{\boldsymbol w) }\ dV  =  \int_{\Omega}\dfrac{\text {Re} }{\Delta t} \boldsymbol u^{n-1} \cdot \boldsymbol w\ dV + \int_{\Gamma _N} \left( -p^n {\boldsymbol n}   + \beta \nabla {\boldsymbol u^n} \cdot {\boldsymbol n} + {\boldsymbol \tau^n }\cdot {\boldsymbol n} \right) \cdot {\boldsymbol w} \ dA,
    \label{eqn:momentum_weak}
    \end{align}
    \begin{align}
    \int_{\Omega} \left( \nabla \cdot {\boldsymbol u^n } \right) q  \ dV = 0,
    \label{eqn:incomp_weak}
    \end{align}
        \begin{align} 
    & \int_{\Omega} \left(1 + \dfrac{\text{We} }{\Delta t}\right) {\boldsymbol \tau^n } : {\boldsymbol \sigma}\ dV  + \int_{\Omega} \text{We} ({\boldsymbol u^n}\cdot \nabla) \boldsymbol \tau^n : {\boldsymbol \sigma} \ dV  \nonumber - \int_{\Omega} [({\nabla \boldsymbol u^n}) {\boldsymbol \tau^n }   +   {\boldsymbol \tau^n } ({\nabla \boldsymbol u^n})^T )]: {\boldsymbol \sigma}  \ dV  \nonumber \\
    & - \int_{\Omega} (1 -\beta)  ({\nabla \boldsymbol u^n} : {\boldsymbol \sigma} + ({\nabla \boldsymbol u^n})^T : {\boldsymbol \sigma})  \ dV = \int_{\Omega} \left(\dfrac{\text{We} }{\Delta t}\right) {\boldsymbol \tau^{n-1} } : {\boldsymbol \sigma} \ dV\,. 
       \label{eqn:const_weak}
    \end{align} 
    \label{eqn:weaks}
    \end{subequations}
We make use of finite element approximations, and discretize in space by partitioning the domain into quadrilaterals (2D) or hexahedra (3D). The velocities are approximated using piecewise continuous biqudratic (2D) or triquadratic (3D) polynomials, denoted by $Q_2$, and the pressures by piecewise discontinuous linear polynomials, denoted by $P_1^{disc}$. This combination of elements satisfies the velocity-pressure inf-sup stability condition \cite{Hughes1987, Turek1999}. The element choice for the extra stress is dependent on further discretization of (\ref{eqn:weaks}). This is addressed in the next section.

\subsection{Discontinuous Galerkin method}

Discontinuous Galerkin methods are a class of finite element methods in which the continuity requirement across elements is relaxed. The method was introduced by Reed and Hill in 1974 \cite{Reed1973} and  by Lesaint and Raviat \cite{Lesaint1974} in 1974, to solve the neutron transport problem. The discontinuous Galerkin method offers various advantages such as the ability to handle complex geometries easily, incorporating refinement which may result in neighbouring elements having differing polynomial orders, or multiple cells sharing the same interface with one element ($hp$-refinement). Furthermore it can be easily parallelized. A more important advantage is that the method is capable of capturing  discontinuous solutions that arise in some hyperbolic problems. Additionally, the method allows for solutions to be determined on an element-by-element basis. 

Discontinuous Galerkin Methods were first used to solve for viscoelastic flows by Fortin and Fortin \cite{Fortin1989}, and more particularly for the Oldroyd-B model in \cite{Boyaval2008, Donev2014}. In these works, and in the current study, upwinding is used to address the instabilities arising from advection-dominated problems, such as that for the extra stress constitutive relation.\\

The element boundaries are split into their upwind and downwind components $\Gamma_+$ and $\Gamma_-$. 
 We denote the jump of a variable $u$ across an inter-element boundary by
    \begin{align*}
    [\![ u ]\!] =  u^+ - u^-,
    \end{align*}
where $u^+$ and $u^-$ are the values of $u$ on the upwind and downward element boundaries respectively. Upwinding is achieved by integrating by parts twice the terms on the right-hand side of \eqref{eqn:const_weak}. This equation becomes, on an element $\Omega_e$ with boundary $\Gamma_e$,
    \begin{align} 
    & \int_{\Omega _e} \left(1 + \dfrac{\text{We} }{\Delta t}\right) {\boldsymbol \tau^n } : {\boldsymbol \sigma}  \ dV  + \int_{\Omega _e}  \text{We}\, (\boldsymbol u^n\cdot{ \nabla \boldsymbol \tau^n }): {\boldsymbol \sigma}  \ dV
    - \int_{\Omega _e} ({\nabla \boldsymbol u^n}  {\boldsymbol \tau^n } + {\boldsymbol \tau^n }  ({\nabla \boldsymbol u^n})^T ): {\boldsymbol \sigma}   \ dV \nonumber \\
    &- \int_{\Omega _e}  (1 -\beta)  ({\nabla \boldsymbol u^n} : {\boldsymbol \sigma} + ({\nabla \boldsymbol u^n})^T : {\boldsymbol \sigma})  \ dV \nonumber \\ &= \int_{\Omega _e} \left(\dfrac{\text{We} }{\Delta t}\right) {\boldsymbol \tau^{n-1} } : {\boldsymbol \sigma}  \ dV - \int_{\Gamma _e} \text{We}\,(\boldsymbol u^{n+} \cdot \boldsymbol n^+) [\![ {\boldsymbol \tau^n } ]\!]: {\boldsymbol \sigma^+}   \ dA\,.
    \label{eqn:const_weak_dg}
    \end{align} 
Here ${\boldsymbol n}^+$ and ${\boldsymbol n}^-$ denote respectively the outward unit normals on the upwind and downwind parts of the boundary. For the extra stress field piecewise constant $Q_0$ or bi- or trilinear discontinuous $Q_1^{\rm disc}$ elements are used. The non-linear governing equations are linearized using a modified Newton-Raphson scheme with damping. The resulting model for the Oldroyd-B fluid was implemented in the c++ Finite Element library deal.ii \cite{Bangerth2007}.

\section{A benchmark problem}

\begin{figure}[H]
\begin{center}
\includegraphics[scale = 0.6]{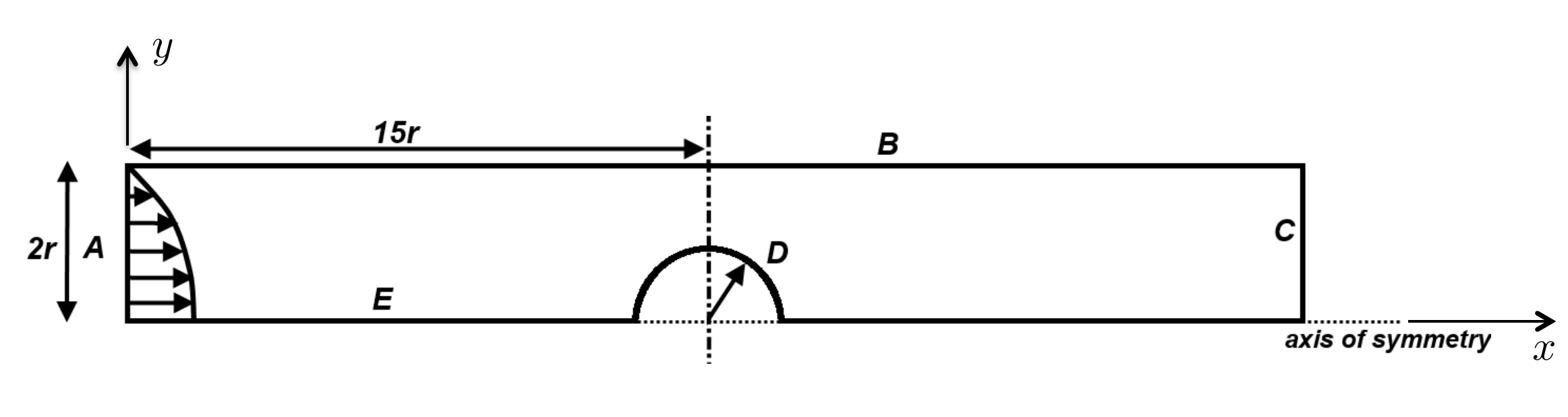}
\caption{Geometry of channel with cylindrical obstruction (D) of radius $r$, with inlet A, channel wall B and outlet C}
\label{img:geo}
\end{center}
\end{figure}

We consider the flow of a fluid in a channel with a cylindrical obstruction. The cylinder is placed symmetrically in the channel, so that it suffices to consider flow in the domain shown in Figure \ref{img:geo}. 
The channel is rectangular with a width $4r$, where $r$ is the radius of the symmetrically placed cylindrical obstruction. The channel half-length $15r$ is sufficiently long for the flow to fully develop, and ensures that the boundary conditions do not affect the behaviour at the obstruction.The geometry of the problem has no singularities, but the challenge lies in predicting the sharp stress boundary layers that arise around the obstruction and along the axis of symmetry in the wake of the obstruction \cite{Afonso2009}.\\

The majority of studies have focused on obtaining solutions for high Weissenberg numbers with different numerical methods. In these studies most schemes fail to converge at Weissenberg numbers of 0.7 - 0.8. The problem has also been shown to be mesh-sensitive with increasing Weissenberg number. Though some solutions have been obtained for higher Weissenberg numbers, there is still uncertainty about the accuracy of the solutions obtained [42]. Nevertheless there is some agreement on the expected behaviour for this benchmark problem.\\


The boundary conditions for the velocity are set to the analytical solution for Poiseuille flow at the inlet, together with a weakly enforced zero extra stress. No-slip boundary conditions are set along walls B and D, and traction-free conditions at the outlet. Along the axis of symmetry the conditions are 
\begin{align*}
\dfrac{\partial \boldsymbol u }{ \partial n} = \boldsymbol 0,\quad \dfrac{\partial \boldsymbol \tau }{ \partial n} =   \boldsymbol 0,\quad
 \boldsymbol u \cdot \boldsymbol n = 0.
\end{align*}
The following parameters are used: Re = 0.1 to approximate creep flow, $\beta = 0.59$ and We ranging from $0.1$ to $0.5$, as in  \cite{Afonso2009} and  \cite{Donev2014}. The ratio of cylinder radius to channel half-width is set at $D/r = 1$. Solutions were obtained on meshes of increasing refinement with adaptive mesh refinement being used after the first global refinement. The mesh properties are summarized in Table \ref{tab:mesh_sizes}.  
\begin{table}[H]
    \centering
    \begin{tabular}{  c  c  c  c }
    \hline
    ~~~~Mesh~~~~ & Number of elements & Mesh parameter $h$ & Total degrees of freedom \\ 
    \hline
    M0 & 502 & 0.208 & 2008 \\  
    M1 & 2008 & 0.0983 & 8230 \\
    M2 & 3214 & 0.0478 & 12856 \\
    M3 & 5143 & 0.0235 & 20572 \\
    M4 & 8230 & 0.0117 & 32590 
    \\ \hline
    \end{tabular}
    \caption{Summary of mesh properties for benchmark problem \\}
    \label{tab:mesh_sizes}
\end{table}

    There is little variation between the meshes M3 and M4 and so results are presented corresponding to the mesh M4. The algorithm converged for ${\rm We}=0.1 - 0.6$ for $Q_0$ elements while for $Q_1^{\rm disc}$ a solution could not be obtained at ${\rm We}=0.6$. Similar limitations have been encountered in other works (see for example  \cite{Afonso2009, Coronado2006, Coronado2007, Hulsen2005,Jensen2015}).\\

\begin{figure}[H]
    \begin{center}          
    \begin{subfigure}{.8\textwidth}
    \centering
    \includegraphics[scale = 0.3]{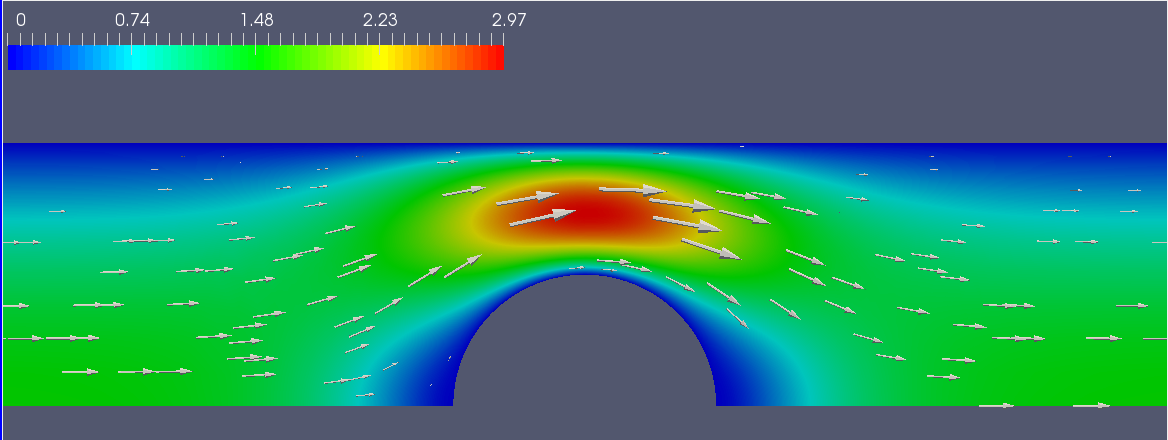}
    \caption{velocity}
    \end{subfigure}
    
    \begin{subfigure}{.8\textwidth}
    \centering
    \includegraphics[scale = 0.3]{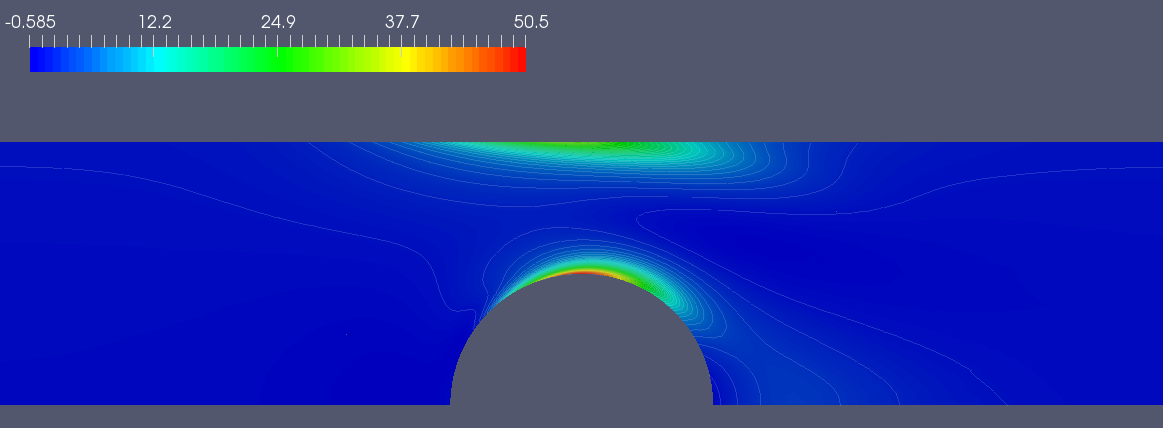}
    \caption{$\tau_{xx}$}
    \end{subfigure}
    
    \begin{subfigure}{.8\textwidth}
    \centering
    \includegraphics[scale = 0.3]{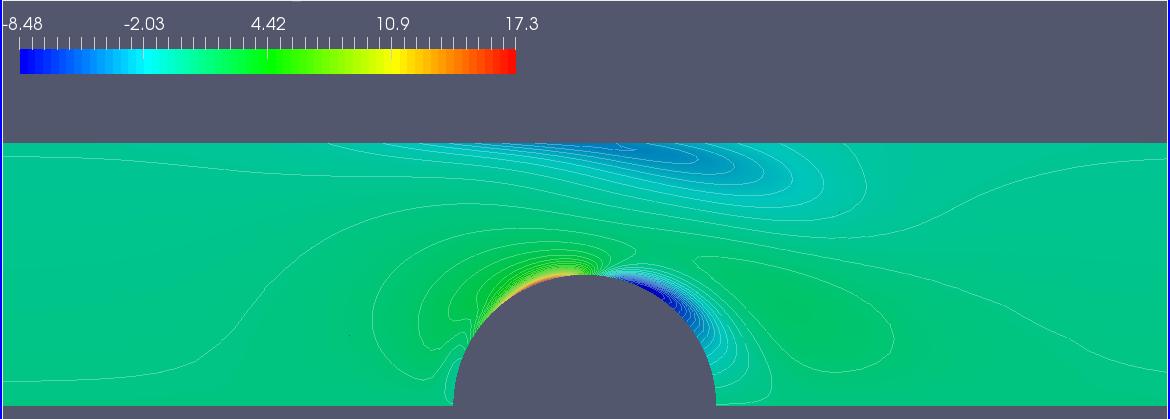}
    \caption{$\tau_{xy}$}
    \end{subfigure}
    
    \begin{subfigure}{.8\textwidth}
    \centering
    \includegraphics[scale = 0.3]{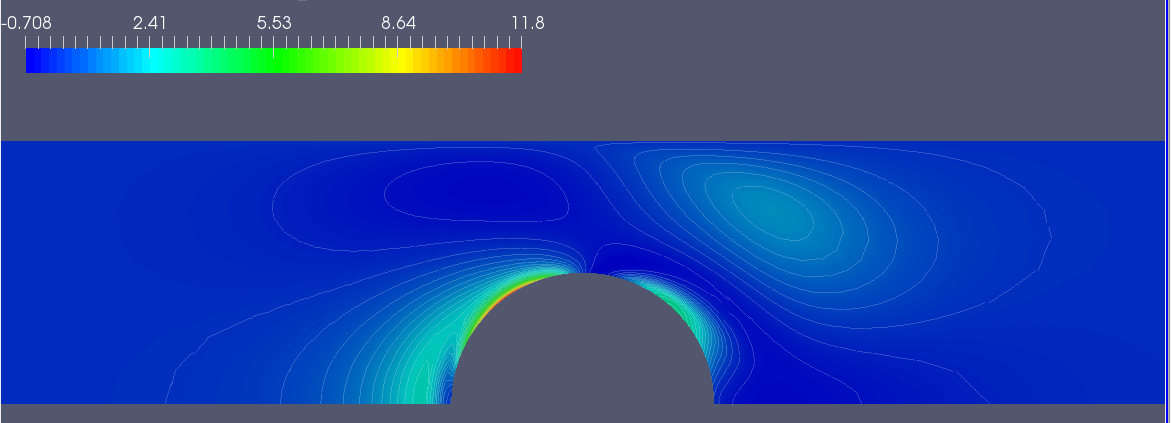}
    \caption{$\tau_{yy}$}
    \end{subfigure}
    
    \caption{Steady state solution for the Oldroyd-B fluid ($\mbox{We} = 0.3$)}
    \label{img:sol_prof}
    \end{center}
\end{figure}

The solution profiles obtained for ${\rm We} = 0.3$ are shown in Figure  \ref{img:sol_prof}.

The velocity profile obtained is very similar  to that for the Newtonian fluid, as seen also in other related studies \cite{Kim2004, Claus2013, Choi2012}. The direct extra stress component $\tau_{xx}$ is positive at the apex of the cylinder and the channel wall directly above it, indicating tensile behaviour. Further stretching is observed in the wake of the cylinder. The shear component $\tau_{xy}$ shows a maximum on the upstream side of the cylinder and a decline on the downstream side. The component $\tau_{yy}$ has a maximum on the left side of the cylinder and a smaller local maximum on the right  side. A stress-free zone is seen in the zero velocity region downstream of the cylinder, with $\tau_{yy}$ having the steepest gradient close to that area. These trends are similar to those observed in \cite{Kim2004, Claus2013, Choi2012}. A plot of the direct stress component along the axis of symmetry and cylinder wall, shown in Figure 3, shows that the extra stress increases with increasing Weissenberg number. 

\begin{figure}[H]
    \begin{center}
    \begin{subfigure}{.5\textwidth}
        \input{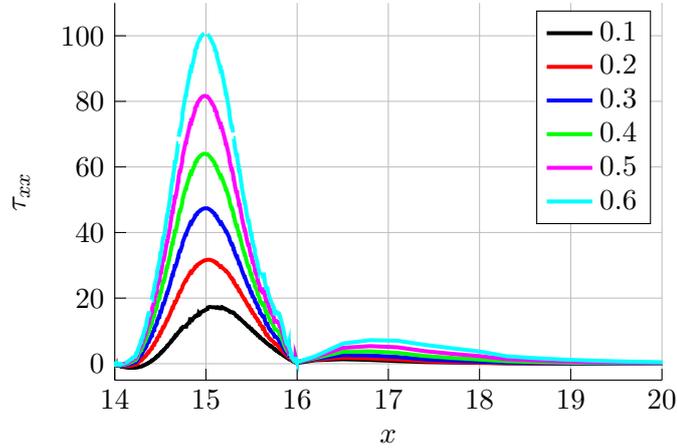}
        \caption{}
        \end{subfigure}
        \begin{subfigure}{.5\textwidth}
    \input{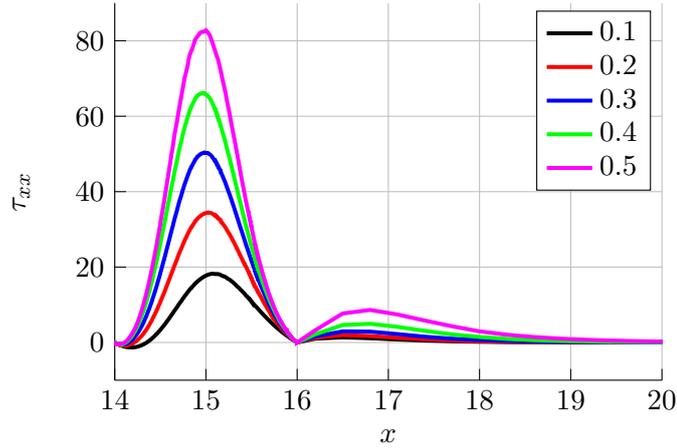}
    \caption{}
    \end{subfigure}
    \caption{Direct stress $\tau_{xx}$ for increasing  Weissenberg number: (a) using $Q_0$ elements, and (b) $Q_1^{\rm disc}$ elements for the extra stress. The centre of the channel and cylinder is located at $x = 15$.} 
    \label{img:var_we}
    \end{center}
\end{figure}

The maximum stress occurs at the apex of cylinder with a much smaller peak in the wake of the cylinder, as seen also in \cite{Donev2014, Fan1999, Kim2004, Hulsen2005, Claus2013, Choi2012},  with differences mainly occurring at higher We ($0.5$ and above) where the maxima are higher and the peak in the wake lower. 

\subsection{Dimensionless drag}

The dimensionless drag $F_D$ over the cylinder (D) is defined by
\begin{equation}
F_D = -2 \int_D \boldsymbol{e}_x \cdot(\underbrace{-p\boldsymbol{I} + \beta(\nabla \boldsymbol{u} + (\nabla \boldsymbol{u})^T) + \boldsymbol{\tau}}_{\boldsymbol T})\,\boldsymbol {n}\ dA
\label{drag}
\end{equation}
where $D$ denotes the surface of the cylinder, $\boldsymbol{e}_x$ is the unit vector in the direction of the axis of the cylinder, and $\boldsymbol n$ is the outward unit normal to this surface. Values of $F_D$ were obtained for both $Q_0$ and $Q_1$ elements at mesh refinement M2, for consistent comparison with the literature. A comparison for varying values of We is presented in Table \ref{tab:drag}.
\\

\begin{table}[H]
\centering
\begin{tabular}{  c c c c c c c c c c }
\hline
We & M2 Q${_0}$ & M2 Q${_1}$  & Donev \cite{Donev2012} & Fan \cite{Fan1999}  & Kim \cite{Kim2004} & Hulsen \cite{Hulsen2005} &  Claus \cite{Claus2013} \\
\hline
0.1 & 129.297 & 130.311 & 130.558 & 130.36 & 130.359 & 130.363 & 130.364 \\
0.2 & 125.546 & 126.511 & 126.629 & 126.62 & 126.622 & 126.626 & 126.626 \\
0.3 & 122.371 & 123.018 & 123.089 & 123.19 & 123.188 & 123.193 & 123.192 \\
0.4 & 120.256 & 120.396 & 120.393 & 120.59 & 120.589 & 120.596 & 120.593 \\
0.5 & 119.237 & 118.660 & 118.656 & 118.83 & 118.824 & 118.836 & 118.826
\\ \hline
\end{tabular}
\caption{Table of dimensionless drag $F_D$ compared to literature.}
\label{tab:drag}
\end{table}

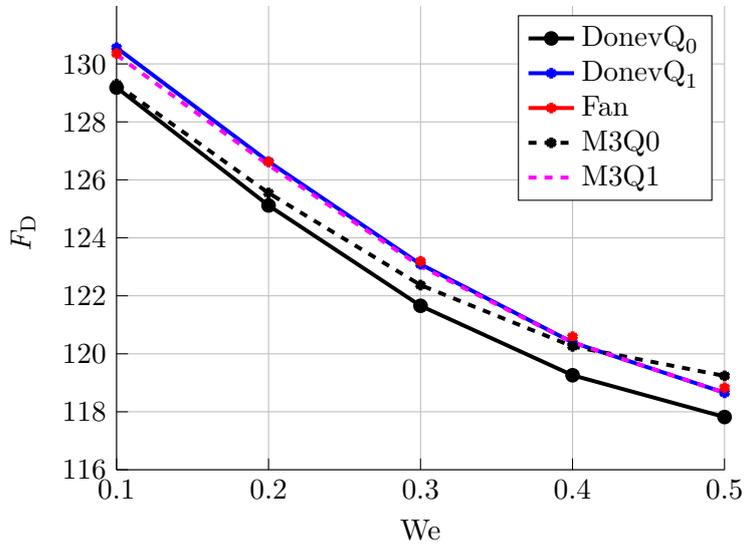
\begin{figure}[H]
\begin{center}
%
%
\definecolor{mycolor1}{rgb}{1.00000,0.00000,1.00000}%
\begin{tikzpicture}

\begin{axis}[%
width=8cm,
height=6.15cm,
at={(0.831in,0.529in)},
scale only axis,
every outer x axis line/.append style={black},
every x tick label/.append style={font=\color{black}},
every x tick/.append style={black},
xmin=0.1,
xmax=0.5,
xlabel={We},
every outer y axis line/.append style={black},
every y tick label/.append style={font=\color{black}},
every y tick/.append style={black},
ymin=116,
ymax=132,
ylabel={$\text{\it{} F}_\text{D}$},
axis background/.style={fill=white},
axis x line*=bottom,
axis y line*=left,
xtick={0.1, 0.2, ..., 0.5},
ytick={116, 118, 120, ..., 130},
xmajorgrids,
ymajorgrids,
legend style={legend cell align=left, align=left, draw=black}
]
\addplot [color=black, line width=1.5pt, mark=*, mark options={solid, black}]
  table[row sep=crcr]{%
0.1	129.183\\
0.2	125.116\\
0.3	121.655\\
0.4	119.26\\
0.5	117.819\\
};
\addlegendentry{$\text{DonevQ}_\text{0}$}

\addplot [color=blue, line width=1.5pt, mark=asterisk, mark options={solid, blue}]
  table[row sep=crcr]{%
0.1	130.558\\
0.2	126.629\\
0.3	123.089\\
0.4	120.393\\
0.5	118.656\\
};
\addlegendentry{$\text{DonevQ}_\text{1}$}

\addplot [color=red, line width=1.5pt, draw=none, mark=asterisk, mark options={solid, red}]
  table[row sep=crcr]{%
0.1	130.36\\
0.2	126.62\\
0.3	123.19\\
0.4	120.59\\
0.5	118.83\\
};
\addlegendentry{Fan}

\addplot [color=black, dashed, line width=1.5pt, mark=asterisk, mark options={solid, black}]
  table[row sep=crcr]{%
0.1	129.297\\
0.2	125.546\\
0.3	122.371\\
0.4	120.256\\
0.5	119.237\\
};
\addlegendentry{M3Q0}

\addplot [color=mycolor1, dashed, line width=1.5pt]
  table[row sep=crcr]{%
0.1	130.311\\
0.2	126.511\\
0.3	123.018\\
0.4	120.396\\
0.5	118.66\\
};
\addlegendentry{M3Q1}

\end{axis}
\end{tikzpicture}%
\caption{Dimensionless drag profile for  Q$_{0}$ and Q$_{1}$ elements compared to \cite{Donev2014, Fan1999}.}
\label{img:drag}
\end{center}
\end{figure}

The values obtained using $Q_1$ elements show close correlation with values in the literature. The values obtained using $Q_0$ elements are slightly less accurate, as observed in \cite{Donev2012}. A comparison of the dimensionless drag with results obtained in \cite{Fan1999, Donev2012} is shown in Figure \ref{img:drag}, using mesh M3.

These results show the degree of improvement in results with the use of piecewise-blinear as opposed to piecewise-constant elements.

\section{Blood flow in an arteriovenous fistula}
An arteriovenous fistula is a mode of vascular access formed by connecting an artery and a vein. Arteriovenous fistulae (AVF) are used in haemodialysis, which is required by most patients with late stage renal disease. For this treatment blood is extracted from the body into a filter through a tube. The process requires blood flow rates above 300 ml/min \cite{Decorato2011}. Computational fluid dynamics simulations on AVFs include the works \cite{Lee2007,Boghosian2014,Botti2013,VanCanneyt2010,Niemann2010}, where flow features such as  recirculation, stagnation and separation are apparent.\\

\begin{figure}[H]
\begin{center}
\includegraphics[scale = 0.4]{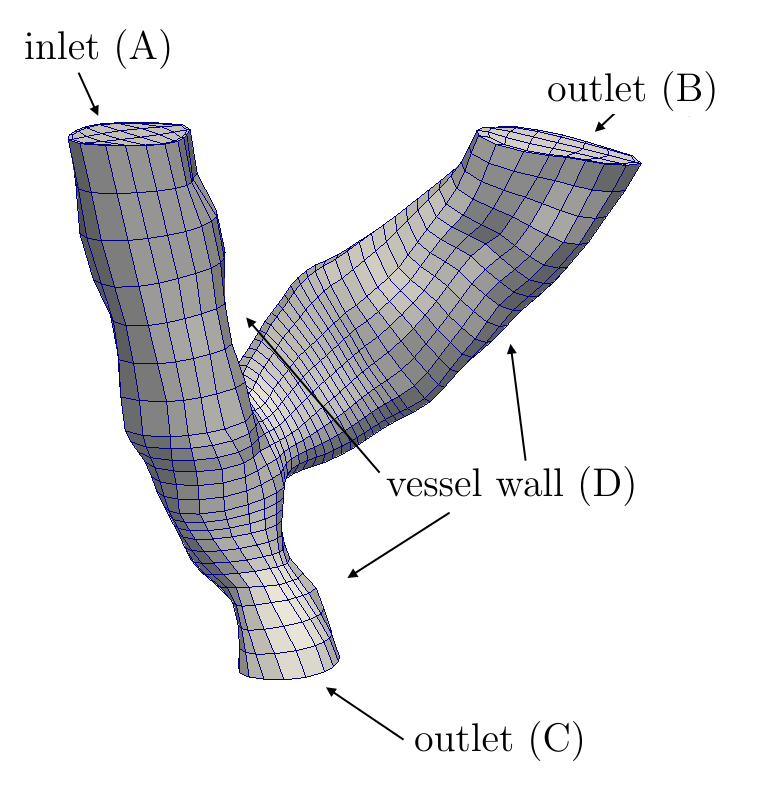}
\caption{Geometry of the arteriovenous fistula processed from MRI data. The artery comprises the region from inlet (A) to outlet (C), and the vein extends from outlet (B)}
\label{img:fistula}
\end{center}
\end{figure}

In this work the aim is to compare results assuming Newtonian behaviour for blood flow in the AVF, with those for viscoelastic fluids. It appears that such a comparison is not available in the current literature. The geometry of the fistula is obtained from a patient-specific geometry extracted from velocity encoded MRI data \cite{DeVilliers2017}. The walls are assumed to be rigid here. The geometry of the fistula is shown in Figure \ref{img:fistula}.
\\

\begin{figure}[H]
    \begin{center}
    \includegraphics[scale = 0.4]{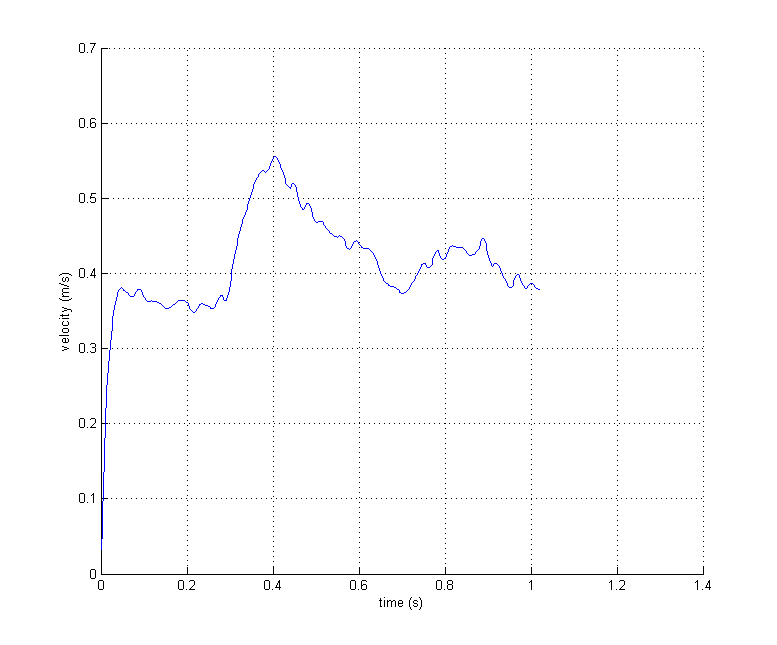}
    \caption{Time history of the maximum velocity at the fistula inlet from velocity encoded MRI scans}
    \label{img:velocity_pulse}
    \end{center}
 \end{figure}

The inlet flow used is based on that obtained from the MRI data. The velocity pulse at the inlet is shown in Figure \ref{img:velocity_pulse}.

The velocity cross-section profile used is a paraboloid of the form
\begin{align}
u_z(x,y) = \left(\dfrac{(x + 0.1865)^2}{0.00405^2} + \dfrac{(y - 0.0137)^2}{0.0048^2}\right) - 1.
    \label{eqn:inlet_para}
\end{align}  
For the outlet boundary condition we take into account the surrounding vascular system by using a resistance boundary condition coupled with backflow stabilization. The resistance boundary condition takes into account pressure wave propagation in the vascular system, while backflow stabilization prevents divergence caused by fluid flowing back into the domain due to the pulsatile nature of the flow \cite{Bazilevs2009, EsmailyMoghadam2011, Vignon2006}. The combined boundary condition is 
\begin{align}
\boldsymbol n\cdot\tilde{\boldsymbol \sigma} \boldsymbol n + R_{\rm out} \int_{\Gamma_{\rm out}}  \boldsymbol u \cdot \boldsymbol n  dA + p_0 = 0,
\label{eqn:resistive_bc}
\end{align}
where 
\begin{align*}
\tilde{\boldsymbol \sigma} \boldsymbol n = -p \boldsymbol n + \eta \boldsymbol D \boldsymbol n - \rho\boldsymbol u (\boldsymbol u \cdot \boldsymbol n)_-\,.
\end{align*}
Here $R_{\rm out}$ is the resistance of the downstream vasculature and $(\boldsymbol u \cdot \boldsymbol n)_-$ is defined by 
\[ 
(\boldsymbol u \cdot \boldsymbol n)_- =  \left\{
\begin{array}{ll}
\boldsymbol u \cdot \boldsymbol n &  ~~~ \text{ if } \boldsymbol u \cdot \boldsymbol n < 0,\\
0 &  ~~~ \text{ if } \boldsymbol u \cdot \boldsymbol n \geq 0.
\end{array} 
\right. 
\]
The parameters used are $p_0 = 85\,$\,{\em mm Hg}, and $R_{\rm out} =  \num{1e4}\,kg / (m^3s)$ and $\num{1e3} ~ kgm^{-3}s^{-1})$ at outlets B and C respectively. 

\subsection{Newtonian simulations}
\begin{center}
    \begin{figure}[H]
    \begin{tabular}{ccc}
    \includegraphics[scale = 0.6]{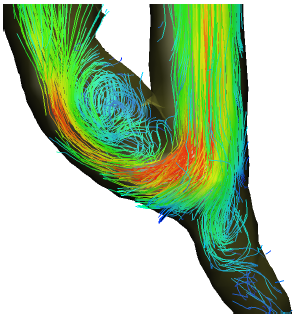}&
    \includegraphics[scale = 0.55]{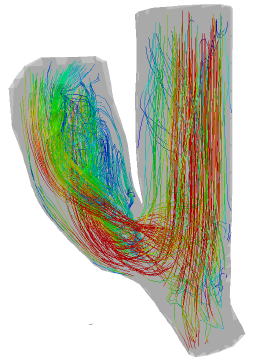}&
    \includegraphics[scale = 0.3]{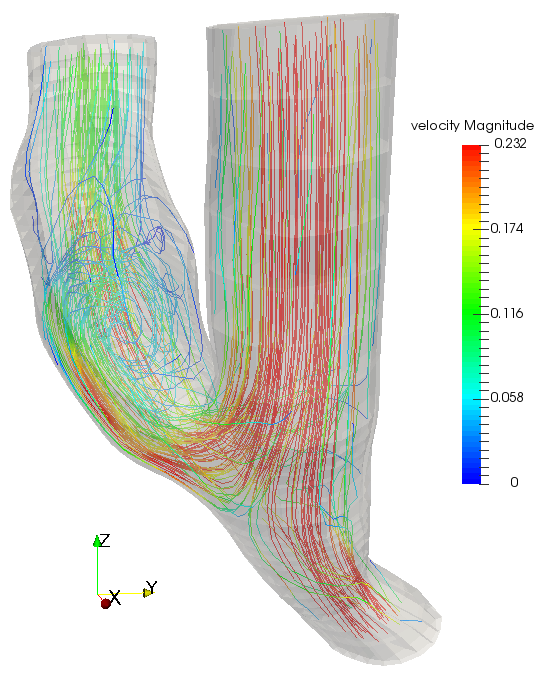}\\
    $t$ = 145ms & & \\
    \includegraphics[scale = 0.6]{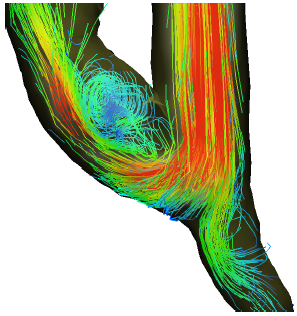}&
    \includegraphics[scale = 0.55]{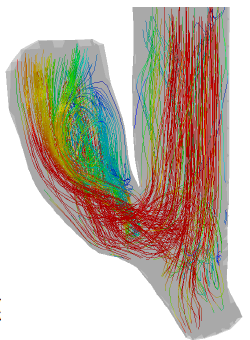}&
    \includegraphics[scale = 0.3]{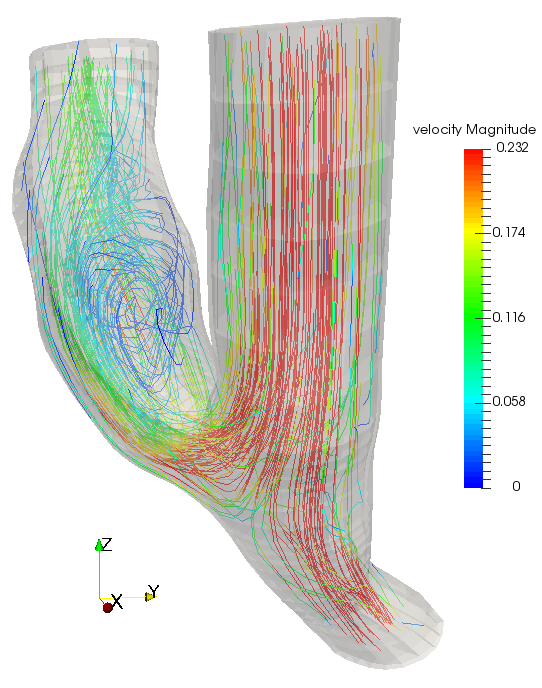}\\
    $t$ = 295ms & & \\
    \includegraphics[scale = 0.6]{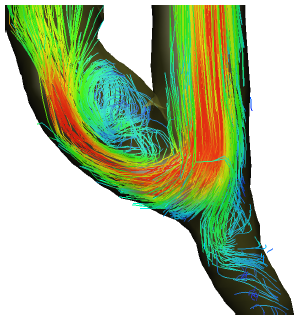}&
    \includegraphics[scale = 0.55]{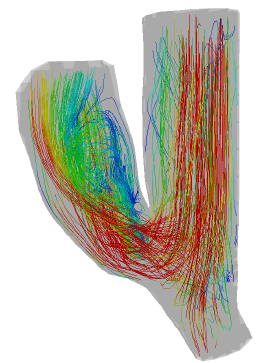}&
    \includegraphics[scale = 0.3]{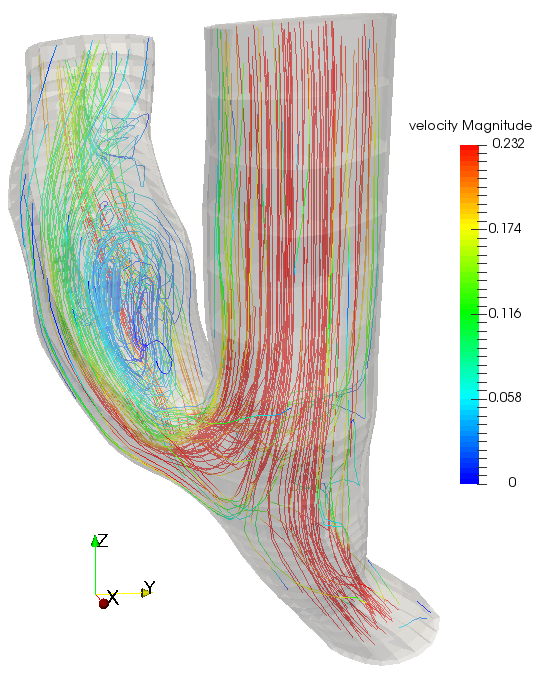}\\
    $t$ = 545ms & & \vspace{3ex}\\
      (a) & (b) & (c)\\
    \end{tabular}
    \caption{Comparison of streamlines: (a) MRI data \cite{DeVilliers2017}; (b) FSI simulations \cite{DeVilliers2017}; (c) current work}
    \label{img:Lfis_newt_mri_andie}
    \end{figure}
    \end{center}

Velocity streamlines are shown in Figure \ref{img:Lfis_newt_mri_andie} for the case of a Newtonian fluid, and compared with the MRI data, obtained together with patient geometry as well as with the results in  \cite{DeVilliers2017} for the deformable domain. The velocities range over 0 to a maximum of 0.232 m/s. There is much similarity across the profiles; this applies in particular to the recirculation region and velocity profiles. 

\begin{figure}[H]
    \begin{center}
    \begin{minipage}[b]{0.45\textwidth}
    \includegraphics[scale = 0.3]{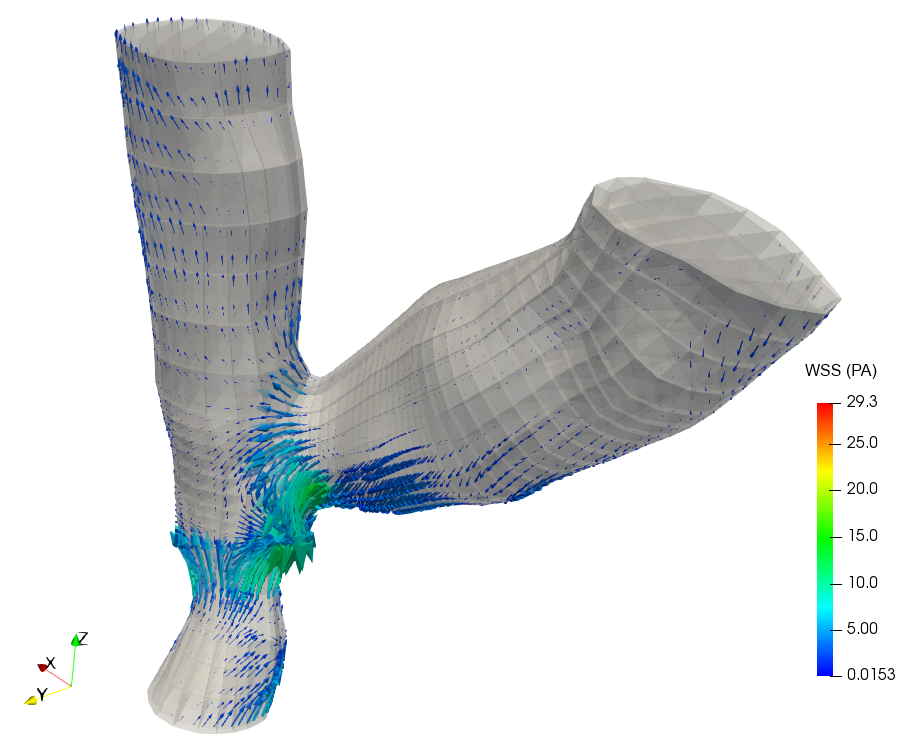}
    \end{minipage}
    \begin{minipage}[b]{0.45\textwidth}
    \includegraphics[scale = 0.45]{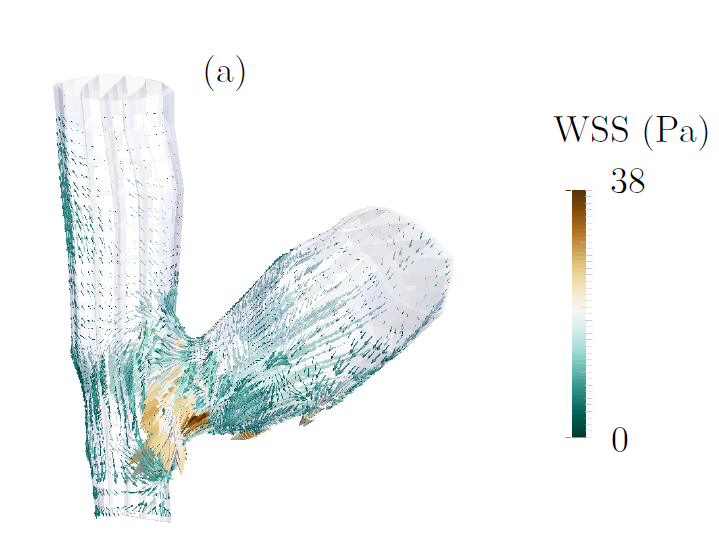}
    \end{minipage}
    \end{center}
    \caption{WSS: (a) current study; (b) simulation in \cite{DeVilliers2017}}
    \label{img:WSS_dim}
    \end{figure}

The maximum wall shear stress WSS, defined by 
\begin{align}
    \text{WSS} = {\boldsymbol \sigma \boldsymbol n} - ({\boldsymbol \sigma \boldsymbol n} \cdot {\boldsymbol n}) \boldsymbol n
    \label{eqn:WSS}
    \end{align}
 is an important parameter as it is much higher in vascular access than in normal physiological conditions. Comparison of the WSS with that in \cite{DeVilliers2017}) at the peak systole is shown in Figure \ref{img:WSS_dim}. The maximum for the present study was found to be equal to 29.7\,Pa (cf. the value of 38Pa found in  \cite{DeVilliers2017}). The difference can be attributed to the rigid wall being used in this work, as well as a difference in the type of resistance boundary condition chosen. The stress profiles are nevertheless similar.

\subsection{Oldroyd-B simulations}
\begin{figure}[H]
    \begin{center}
    \includegraphics[scale = 0.45]{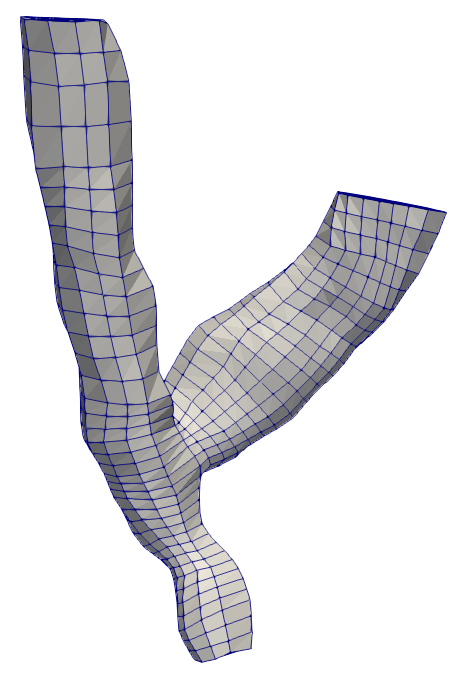}
    \caption{Mesh of the fistula for the case of an Oldroyd-B fluid}
    \label{img:coarse_mesh}
    \end{center}
    \end{figure}

\begin{figure}[H]
    \begin{center}
    \begin{minipage}[b]{0.3\textwidth}
    \includegraphics[scale = 0.4]{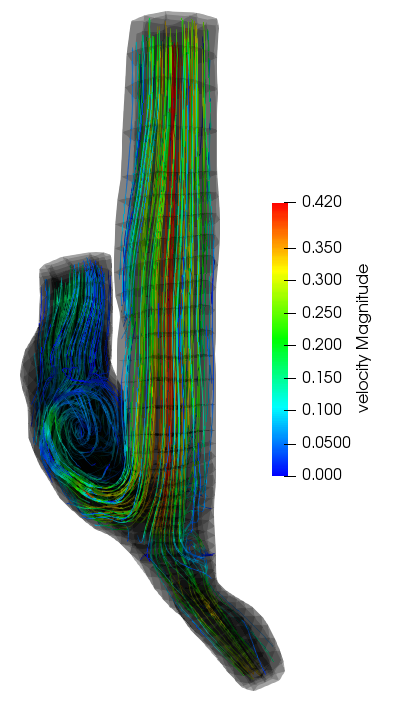}
    \caption*{$t$=145ms}
    \end{minipage}
\hspace{18ex}
    \begin{minipage}[b]{0.3\textwidth}
    \includegraphics[scale = 0.4]{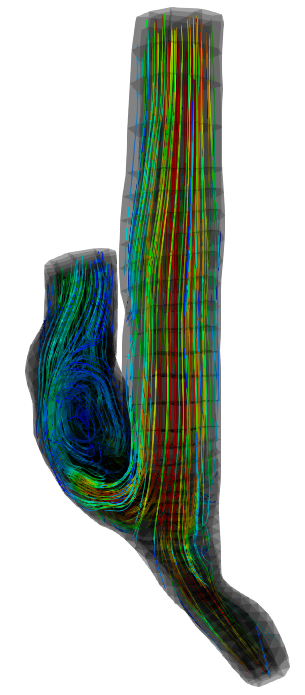}
    \caption*{$t$=545ms}
    \end{minipage}
    \end{center}
    \caption{Velocity streamlines for the Oldroyd-B fluid at We = 0.1}
        \label{velWe01}
\end{figure}

 The simulations for the Oldroyd-B fluid were carried out on a coarser mesh than that used for the Newtonian fluid as the fourfold increase in the degrees of freedom results in a considerable increase in the size of the problem. The mesh used is shown in Figure \ref{img:coarse_mesh}.  The simulations were carried out for ${\rm We} = 0.1$ to $0.5$, which corresponds to the range of relevance for blood \cite{Thurston2006EffectsOF}. Velocity profiles are shown in Figures \ref{velWe01} - \ref{velNewt} for a range of values of We and for the case of a Newtonian fluid. These show similar behaviour to that for the case of a Newtonian fluid, in particular, with a similar recirculation region. However, for ${\rm We}= 0.5$ the velocities are somewhat lower than those at lower values of We, beyond $t=445$ms.

\begin{figure}[H]
    \begin{center}
    \begin{minipage}[b]{0.3\textwidth}
    \includegraphics[scale = 0.4]{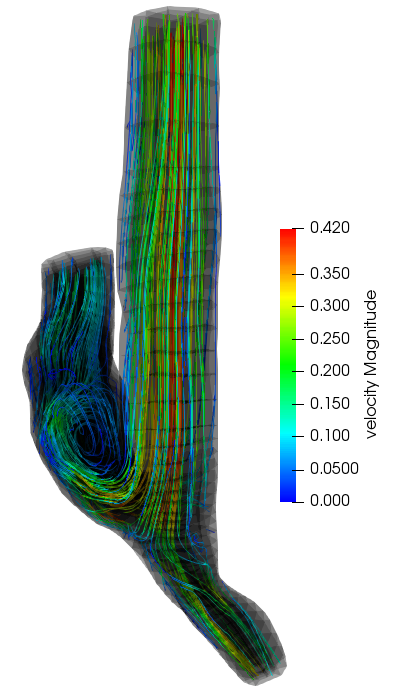}
   \caption*{$t$=145ms}
    \end{minipage}
\hspace{18ex} 
    \begin{minipage}[b]{0.3\textwidth}
    \includegraphics[scale = 0.4]{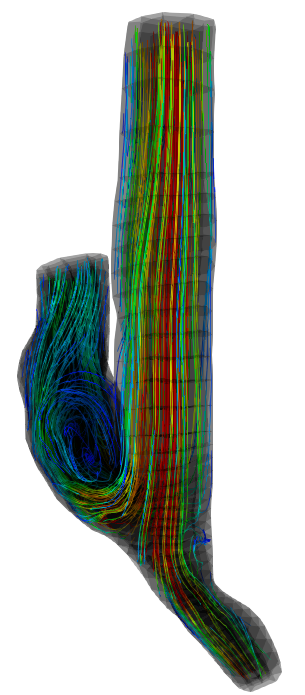}
    \caption*{$t$=545ms}
    \end{minipage}
    \end{center}
    \caption{Velocity streamlines for the Oldroyd-B fluid at We = 0.3}
\end{figure}
    
    
\begin{figure}[H]
    \begin{center}
    \begin{minipage}[b]{0.3\textwidth}
    \includegraphics[scale = 0.4]{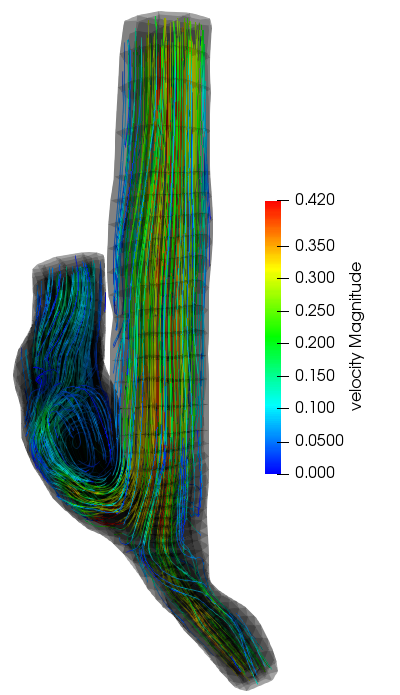}
    \caption*{$t$=145ms}
    \end{minipage}
\hspace{18ex}
    \begin{minipage}[b]{0.3\textwidth}
    \includegraphics[scale = 0.4]{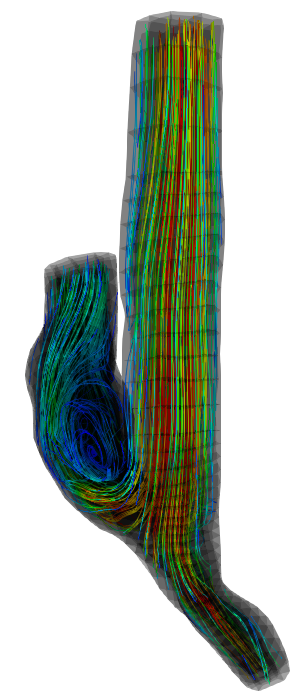}
    \caption*{$t$=545ms}
    \end{minipage}
    \end{center}
    \caption{Velocity streamlines for the Oldroyd-B fluid at We = 0.5}
\end{figure}

\begin{figure}[H]
    \begin{center}
    \begin{minipage}[b]{0.3\textwidth}
    \includegraphics[scale = 0.4]{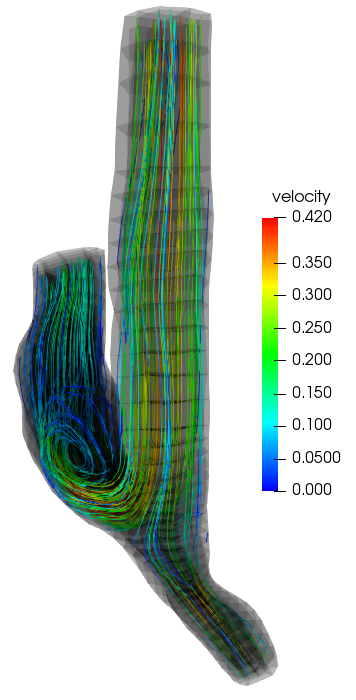}
    \caption*{t=145ms}
    \end{minipage}
\hspace{18ex}
    \begin{minipage}[b]{0.3\textwidth}
    \includegraphics[scale = 0.4]{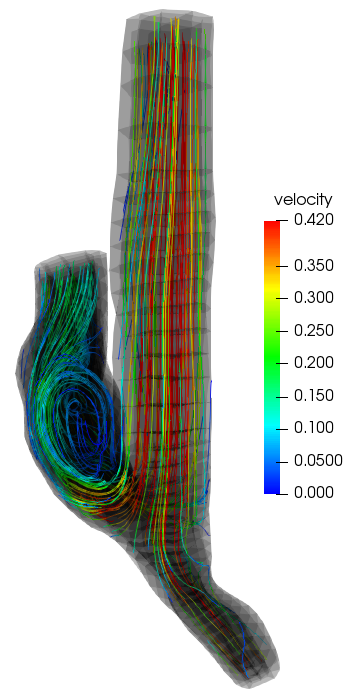}
    \caption*{t=545ms}
    \end{minipage}
    \end{center}
    \caption{Velocity streamlines for the Newtonian fluid}
    \label{velNewt}
\end{figure}
\begin{figure}[H]
    \begin{center}
    \begin{tabular}{cc}
    \includegraphics[scale = 0.4]{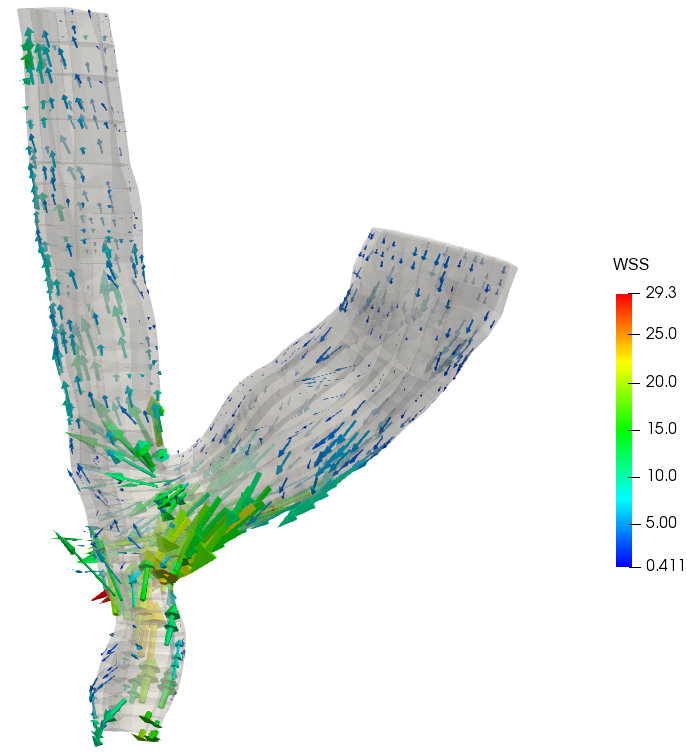} 
    &     \includegraphics[scale = 0.4]{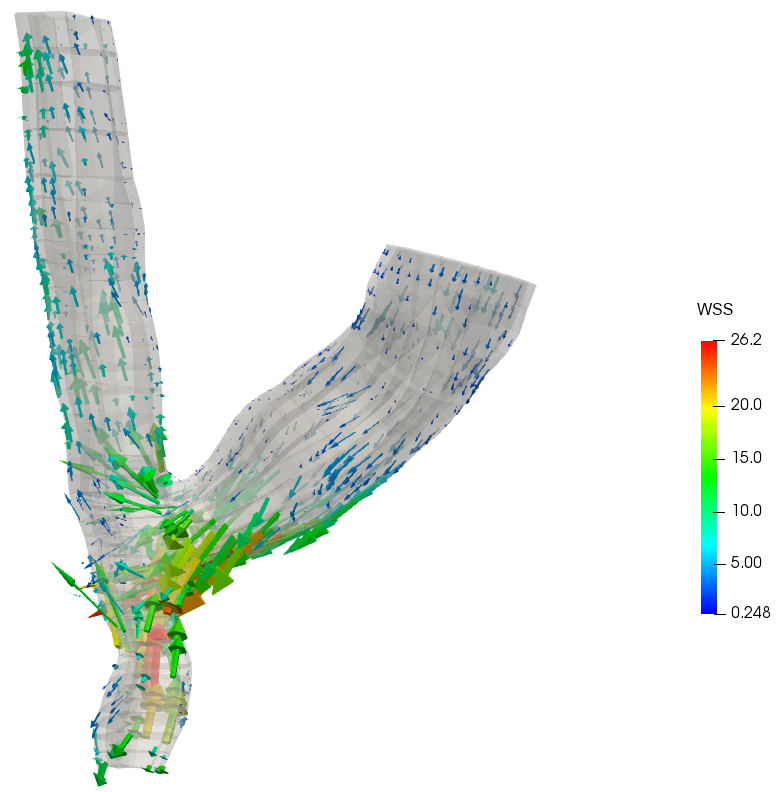} \\
    (a) We = 0.1  & (b) We = 0.3 \\
    \\
    \includegraphics[scale = 0.4]{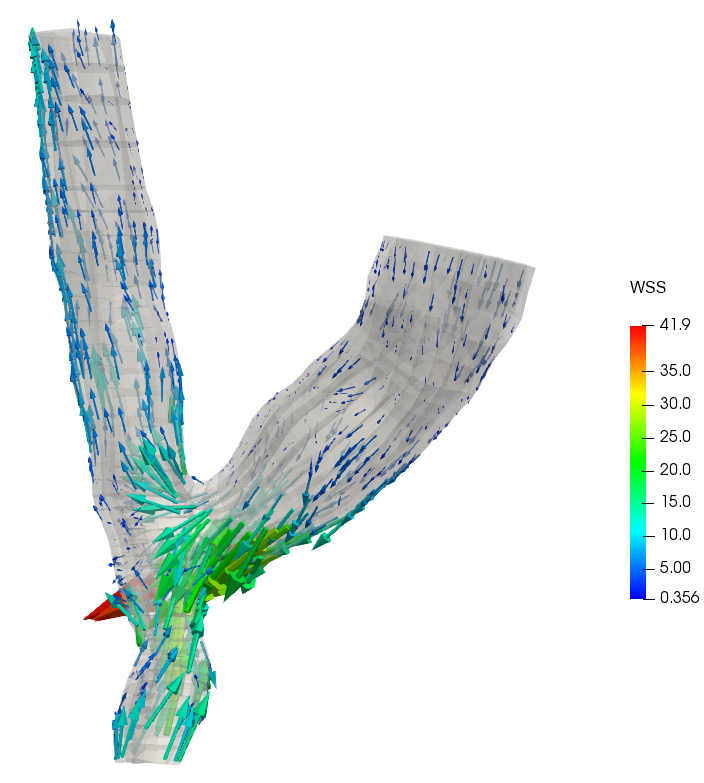} 
    &
   \includegraphics[scale = 0.5]{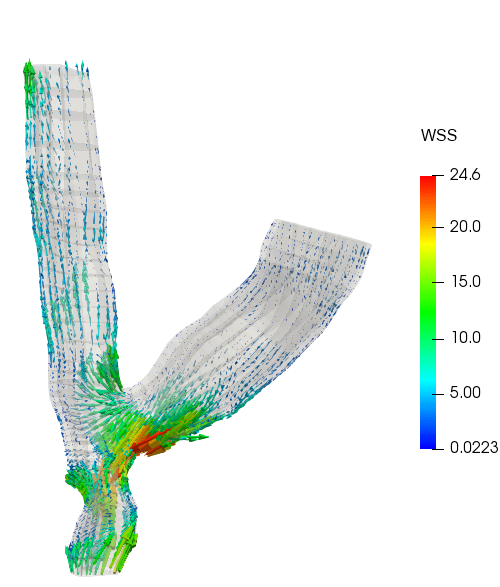} \\
    (c) We = 0.5 & (d) Newtonian Fluid \\
    \end{tabular}
    \end{center}
    \caption{(a) - (c) WSS for varying We; (d) for a Newtonian fluid}
    \label{img:WSS_Visc}
    \end{figure}
\begin{figure}[H]
    \begin{center}
    \includegraphics[scale = 0.5]{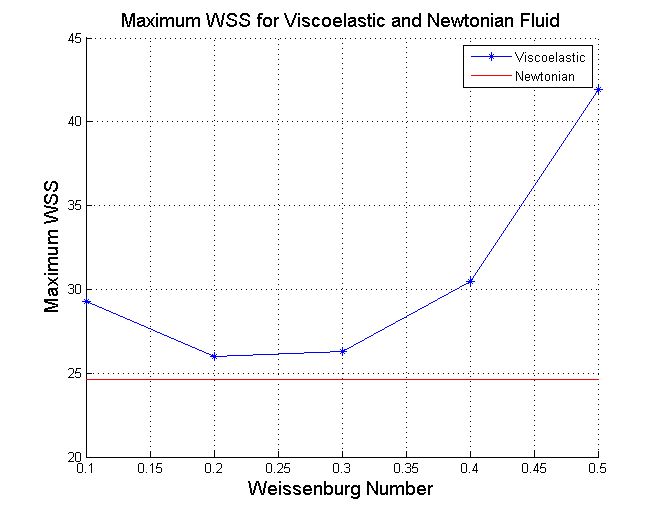}
    \caption{Variation of maximum WSS with Weissenberg number}
    \label{img:wss_plot}
    \end{center}
    \end{figure}

Plots of wall shear stresses are shown in Figure \ref{img:WSS_Visc}. A region of high stress in the vicinity of the junction is evident in all cases. It is also seen that the region of high stress for the Oldroyd-B fluid coincides with that for a Newtonian fluid.  The variation of the WSS with increasing Weissenberg number, shown in  Figure \ref{img:wss_plot}, was found to be roughly parabolic. This is similar to the relation between We and dimensionless drag \cite{Donev2014, Fan1999, Kim2004, Hulsen2005, Claus2013}, with the minimum occurring at a lower value of We for the WSS than for the dimensionless drag.

\section{Conclusions}

This work has concerned a finite element-discontinuous Galerkin analysis of flows of an Oldroyd-B fluid. A two-dimensional benchmark problem has served to ensure appropriate levels of accuracy. The results of the benchmark model are comparable to those in the literature, with differences occurring at higher Weissenberg numbers. The use of piecewise constant approximations for the extra stress suffices to capture adequately the relevant trends in behaviour.\\

Simulations have been carried out of flow of an Oldroyd-B fluid for the complex three-dimensional geometry of a patient-specific arteriovenous fistula, with the assumption of rigid walls. The inlet velocity profile was based on that obtained from MRI data. The flow profiles obtained were similar to those reported for a Newtonian fluid with deformable blood vessels in \cite{DeVilliers2017}. Also, the WSS for the Oldroyd-B fluid shows a similar profile to that of a Newtonian fluid, the difference being a higher stress behind the artery-vein junction for the viscoelastic model. The maximum values of WSS have a strong dependence on Weissenberg number, and are in all cases higher than for a Newtonian fluid. \\

Both the similarities and the differences in behaviour between Oldroyd-B and Newtonian fluids are valuable in determining the most appropriate constitutive models for simulations of the kind considered in this work. Similar remarks apply to the comparisons between results for vessel walls assumed rigid, on the one hand, and deformable on the other.

\bibliographystyle{plain}
\bibliography{myrefs}

\begin{thebibliography}{10}

\bibitem{Afonso2009}
A.~Afonso, P.~J. Oliveira, F.~T. Pinho, and M.~A. Alves.
\newblock {The log-conformation tensor approach in the finite-volume method
  framework}.
\newblock {\em Journal of Non-Newtonian Fluid Mechanics}, 157(1-2):55--65,
  2009.

\bibitem{Bangerth2007}
W.~Bangerth, R.~Hartmann, and G.~Kanschat.
\newblock {deal.II---A general-purpose object-oriented finite element library}.
\newblock {\em ACM Transactions on Mathematical Software}, 33:24/1--24/7, 2007.

\bibitem{Bazilevs2009}
Y.~Bazilevs, J.R. Gohean, T.J.R. Hughes, R.D. Moser, and Y.~Zhang.
\newblock {Patient-specific isogeometric fluid-structure interaction analysis
  of thoracic aortic blood flow due to implantation of the Jarvik 2000 left
  ventricular assist device}.
\newblock {\em Computer Methods in Applied Mechanics and Engineering},
  198(45-46):3534--3550, 2009.

\bibitem{Boghosian2014}
M.~Boghosian, K.~Cassel, M.~Hammes, B.~Funaki, S.~Kim, X.~Qian, X.~Wang,
  P.~Dhar, and J.~Hines.
\newblock {Hemodynamics in the cephalic arch of a brachiocephalic fistula}.
\newblock {\em Medical Engineering and Physics}, 36(7):822--830, 2014.

\bibitem{Botti2013}
L.~Botti, K.~{Van Canneyt}, R.~Kaminsky, T.~Claessens, R.N. Planken,
  P.~Verdonck, A.~Remuzzi, and L.~Antiga.
\newblock {Numerical evaluation and experimental validation of pressure drops
  across a patient-specific model of vascular access for hemodialysis}.
\newblock {\em Cardiovascular Engineering and Technology}, 4(4):485--499, 2013.

\bibitem{Boyaval2008}
S.~Boyaval, L.~Tony, and C.~Mangoubi.
\newblock Free-energy-dissipative schemes for the {Oldroyd-B model}.
\newblock {\em ESAIM: Mathematical Modelling and Numerical Analysis},
  43(3):523--561, 2009.

\bibitem{Choi2012}
Y.J. Choi, M.A. Hulsen, and H.E.H. Meijer.
\newblock {Simulation of the flow of a viscoelastic fluid around a stationary
  cylinder using an extended finite element method}.
\newblock {\em Computers and Fluids}, 57:183--194, 2012.

\bibitem{Claus2013}
S.~Claus and T.~N. Phillips.
\newblock {Viscoelastic flow around a confined cylinder using spectral/hp
  element methods}.
\newblock {\em Journal of Non-Newtonian Fluid Mechanics}, 200:131--146, 2013.

\bibitem{Coronado2007}
O.M. Coronado, D.~Arora, M.~Behr, and M.~Pasquali.
\newblock {A simple method for simulating general viscoelastic fluid flows with
  an alternate log-conformation formulation}.
\newblock {\em Journal of Non-Newtonian Fluid Mechanics}, 147(3):189--199,
  2007.

\bibitem{Coronado2006}
Oscar~M. Coronado, Dhruv Arora, Marek Behr, and Matteo Pasquali.
\newblock {Four-field Galerkin/least-squares formulation for viscoelastic
  fluids}.
\newblock {\em Journal of Non-Newtonian Fluid Mechanics}, 140(1-3):132--144,
  2006.

\bibitem{DeVilliers2017}
A.~M. de~Villiers, A.~T. McBride, B.~D. Reddy, T.~Franz, and B.~S.
  Spottiswoode.
\newblock {A validated patient-specific FSI model for vascular access in
  haemodialysis}.
\newblock {\em Biomechanics and Modeling in Mechanobiology}, 17(2):479--497,
  2017.

\bibitem{Decorato2011}
I.~Decorato, Z.~Kharboutly, C.~Legallais, and A.~V. Salsac.
\newblock {Numerical study of the influence of wall compliance on the
  haemodynamics in a patient-specific arteriovenous fistula}.
\newblock {\em Computer Methods in Biomechanics and Biomedical Engineering},
  14:121--123, 2011.

\bibitem{Donev2012}
I.~Donev.
\newblock Time dependent finite element simulations of a {Generalized Oldroyd-B
  Fluid}.
\newblock Master's thesis, University of Cape Town, 2012.

\bibitem{Donev2014}
I.~G. Donev and B.~D. Reddy.
\newblock {Time-dependent finite element simulations of a shear-thinning
  viscoelastic fluid with application to blood flow}.
\newblock {\em International Journal for Numerical Methods in Fluids},
  75(9):668--686, 2014.

\bibitem{Fan1999}
Y.~Fan, R.~I. Tanner, and N.~Phan-Thien.
\newblock {Galerkin/least-square finite-element methods for steady viscoelastic
  flows}.
\newblock {\em Journal of Non-Newtonian Fluid Mechanics}, 84(2-3):233--256,
  1999.

\bibitem{Vignon2006}
C.~A. Figueroa, I.~E. Vignon-Clementel, K.~E. Jansen, T.~J.~R. Hughes, and
  C.~A. Taylor.
\newblock A coupled momentum method for modeling blood flow in
  three-dimensional deformable arteries.
\newblock {\em Computer Methods in Applied Mechanics and Engineering},
  195:5685--5706, 2006.

\bibitem{Fortin1989}
M.~Fortin and A.~Fortin.
\newblock {A new approach for the FEM simulation of viscoelastic flows}.
\newblock {\em Journal of Non-Newtonian Fluid Mechanics}, 32(3):295--310, 1989.

\bibitem{Guess2016}
W.P. Guess, B.D. Reddy, A.~McBride, B.~Spottiswoode, J.~Downs, and T.~Franz.
\newblock Fluid-structure interaction modelling and stabilisation of a
  patient-specific arteriovenous access fistula.
\newblock {\em https://arxiv.org/abs/1704.07753}, 2017.

\bibitem{Hughes1987}
T.~J.~R. Hughes.
\newblock {\em The Finite Element Method: Linear Static and Dynamic Finite
  Element Analysis}.
\newblock Prentice-Hall, 1987.

\bibitem{Hulsen2005}
Martien~A. Hulsen, Raanan Fattal, and Raz Kupferman.
\newblock {Flow of viscoelastic fluids past a cylinder at high Weissenberg
  number: Stabilized simulations using matrix logarithms}.
\newblock {\em Journal of Non-Newtonian Fluid Mechanics}, 127(1):27--39, 2005.

\bibitem{Jensen2015}
K.E. Jensen, P.~Szabo, and F.~Okkels.
\newblock {Implementation of the log-conformation formulation for
  two-dimensional viscoelastic flow}.
\newblock {\em https://arxiv.org/abs/1508.01041}, 2015.

\bibitem{Kim2004}
Ju~Min Kim, Chongyoup Kim, Kyung~Hyun Ahn, and Seung~Jong Lee.
\newblock {An efficient iterative solver and high-resolution computations of
  the Oldroyd-B fluid flow past a confined cylinder}.
\newblock {\em Journal of Non-Newtonian Fluid Mechanics}, 123(2-3):161--173,
  2004.

\bibitem{Lee2007}
Sang~Wook Lee, David~S. Smith, Francis Loth, Paul~F. Fischer, and Hisham~S.
  Bassiouny.
\newblock {Importance of flow division on transition to turbulence within an
  arteriovenous graft}.
\newblock {\em Journal of Biomechanics}, 40(5):981--992, 2007.

\bibitem{Lesaint1974}
P.~Lesaint and P.-A. Raviart.
\newblock {On a finite element method for solving the neutron transport
  equation}.
\newblock {\em Mathematical Aspects of Finite Elements in Partial Differential
  Equations}, pages 89--145, 1974.

\bibitem{EsmailyMoghadam2011}
E.M. Mahdi, Y.~Bazilevs, T.Y. Hsia, I.E. Vignon-Clementel, and A.L. Marsden.
\newblock {A comparison of outlet boundary treatments for prevention of
  backflow divergence with relevance to blood flow simulations}.
\newblock {\em Computational Mechanics}, 48(3):277--291, 2011.

\bibitem{Niemann2010}
A.~K. Niemann, J.~Udesen, S.~Thrysoe, J.~V. Nygaard, E.~T. Fr{\"{u}}nd, S.~E.
  Petersen, and J.~M. Hasenkam.
\newblock {Can sites prone to flow induced vascular complications in a-v
  fistulas be assessed using computational fluid dynamics?}
\newblock {\em Journal of Biomechanics}, 43(10):2002--2009, 2010.

\bibitem{Owens-Phillips2002}
R.~G. Owens and T.~N. Phillips.
\newblock {\em Computational Rheology}.
\newblock Imperial College Press, 2002.

\bibitem{Reed1973}
W.~H. Reed and T.~R. Hill.
\newblock {Triangular mesh methods for the neutron transport equation}.
\newblock {\em Los Alamos Report LA-UR-73-479}, (836), 1973.

\bibitem{Thurston2006EffectsOF}
George~B. Thurston and Nancy~M. Henderson.
\newblock Effects of flow geometry on blood viscoelasticity.
\newblock {\em Biorheology}, 43 6:729--46, 2006.

\bibitem{Turek1999}
S.~Turek.
\newblock {\em {Efficient Solvers for Incompressible Flow Problems: An
  Algorithmic and Computational Approach}}.
\newblock Springer New York, 1999.

\bibitem{VanCanneyt2010}
Koen {Van Canneyt}, Thierry Pourchez, Sunny Eloot, Caroline Guillame, Alexandre
  Bonnet, Patrick Segers, and Pascal Verdonck.
\newblock {Hemodynamic impact of anastomosis size and angle in side-to-end
  arteriovenous fistulae: A computer analysis}.
\newblock {\em Journal of Vascular Access}, 11(1):52--58, 2010.

\bibitem{Venkatesan2017}
Jagannath Venkatesan and Sashikumaar Ganesan.
\newblock {A three-field local projection stabilized formulation for
  computations of Oldroyd-B viscoelastic fluid flows}.
\newblock {\em Journal of Non-Newtonian Fluid Mechanics}, 247:90--106, 2017.

\bibitem{Vignon-Clementel2010}
I.E. Vignon-Clementel, C.A. Figueroa, K.E. Jansen, and C.A. Taylor.
\newblock {Outflow boundary conditions for 3D simulations of non-periodic blood
  flow and pressure fields in deformable arteries}.
\newblock {\em Computer Methods in Biomechanics and Biomedical Engineering},
  13(5):625--640, 2010.

\bibitem{YELESWARAPU1998257}
K~K Yeleswarapu, M~V Kameneva, K~R Rajagopal, and J~F Antaki.
\newblock {The flow of blood in tubes: theory and experiment}.
\newblock {\em Mechanics Research Communications}, 25(3):257--262, 1998.

\end{thebibliography}
\end{document}